\documentclass[runningheads]{llncs}

\usepackage[T1]{fontenc}
\usepackage{graphicx}
\graphicspath{ {./images/} }
\usepackage{caption}
\usepackage{subcaption}
\usepackage{comment}
\usepackage{hhline}

\usepackage{soul}
\usepackage{amsmath}
\usepackage{amssymb}
\usepackage{mathtools}
\usepackage{graphicx}
\usepackage[dvipsnames]{xcolor}

\DeclareMathOperator*{\argmin}{argmin}

\begin{document}
%
% \title{An Automatic System for Umbilical Artery Doppler Waveform Guidance and Quality Assessment in Ultrasound Images}
% \title{Guiding Operator Toward a Better Umbilical Artery Doppler Image}
\title{An Automatic Guidance and Quality Assessment System for Doppler Imaging of Umbilical Artery}
\titlerunning{An Automatic Guidance and QA System for Doppler Imaging of UA}
\author{First Author\inst{1}\orcidID{0000-1111-2222-3333} \and
Second Author\inst{2,3}\orcidID{1111-2222-3333-4444} \and
Third Author\inst{3}\orcidID{2222--3333-4444-5555}}
\author{Chun Kit Wong\inst{1} \and
Manxi Lin\inst{1} \and
Alberto Raheli\inst{1} \and
Zahra Bashir\inst{2} \and
Morten Bo Søndergaard Svendsen\inst{2} \and
Martin Grønnebæk Tolsgaard\inst{2} \and
Aasa Feragen\inst{1} \and
Anders Nymark Christensen\inst{1}}
\authorrunning{Wong et al.}
\institute{Technical University of Denmark \and
CAMES Rigshospitalet \\
\email{ckwo@dtu.dk}}

% \author{Anonymous authors\inst{1,2}}
% \authorrunning{Anonymous et al.}
% \institute{Anonymous institute 1 \and
% Anonymous institute 2 \\
% \email{****@***.**}}

%\institute{Princeton University, Princeton NJ 08544, USA \and
%Springer Heidelberg, Tiergartenstr. 17, 69121 Heidelberg, Germany
% \institute{***}
\maketitle              % typeset the header of the contribution
\begin{abstract}
% In fetal ultrasound screening, Doppler images on the umbilical artery (UA) are important for monitoring blood supply through the umbilical cord. However, to capture UA Doppler images, a number of steps need to be done correctly: placing the gate at a proper location in the ultrasound image to obtain blood flow waveforms, and judging the Doppler waveform quality. Both of these rely on the operator's experience. The shortage of experienced sonographers thus creates a demand for machine assistance. We propose an automatic system to fill this gap. Using a modified Faster R-CNN we obtain an algorithm that suggests Doppler flow gate locations. We subsequently assess the Doppler waveform quality. We validate the proposed system on 657 scans from a national ultrasound screening database. The experimental results demonstrate that our system is useful in guiding operators for  Doppler image capture and quality assessment.
Examination of the umbilical artery with Doppler ultrasonography is performed to investigate blood supply to the fetus through the umbilical cord, which is vital for the monitoring of fetal health. Such examination involves several steps that must be performed correctly: identifying suitable sites on the umbilical artery for the measurement, acquiring the blood flow curve in the form of a Doppler spectrum, and ensuring compliance to a set of quality standards. These steps rely heavily on the operator's skill, and the shortage of experienced sonographers has thus created a demand for machine assistance. In this work, we propose an automatic system to fill the gap. By using a modified Faster R-CNN network, we obtain an algorithm that can suggest locations suitable for Doppler measurement. Meanwhile, we have also developed a method for assessment of the Doppler spectrum's quality. The proposed system is validated on 657 images from a national ultrasound screening database, with results demonstrating its potential as a guidance system.
\end{abstract}
\begin{figure}[h!]
    \centering

        \includegraphics[width=\textwidth]{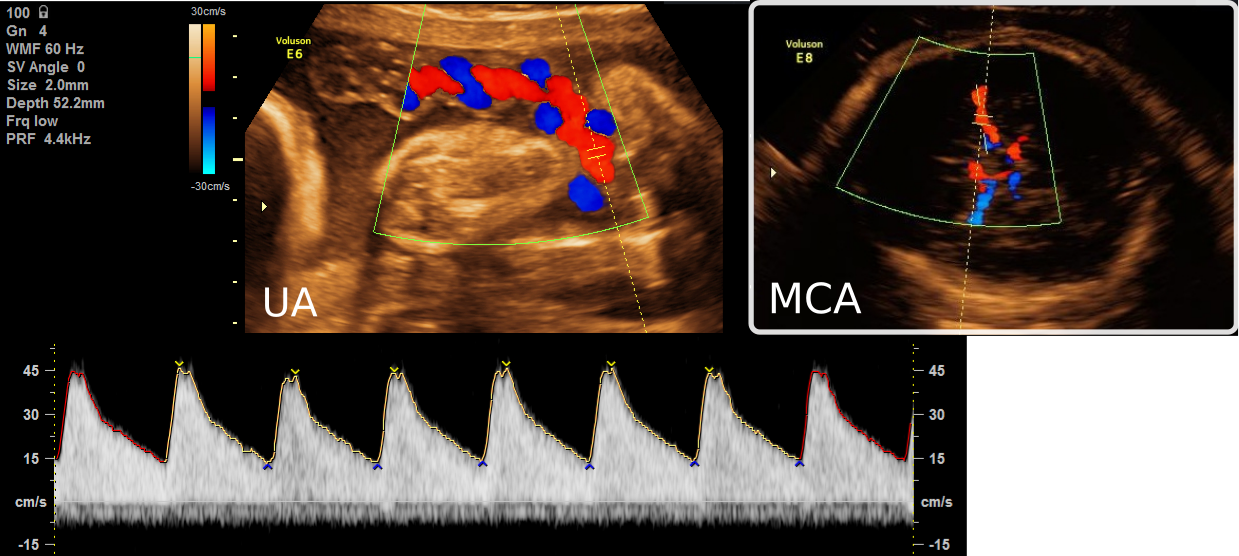}
        \caption{Examples of two fetal standard planes in Doppler ultrasound. \textbf{Left:} Umbilical Artery (UA) including the spectral Doppler waveform. \textbf{Right:} Middle Cerebral Artery (MCA). Note how on the MCA, the vessel is running parallel to the long axis of the vessel's segmentation, making angle estimation from segmentation straightforward. In UA, however, the vessels are usually intertwined, which means the angle cannot be trivially inferred from the segmentation.}
        \label{fig:mca_ua}
\end{figure}

\section{Introduction}
Due to its non-invasive nature, ultrasound is widely used for routine monitoring of fetal health. In the third-trimester, ultrasound screenings are utilized to monitor potential growth restrictions of the fetus and the blood supply through the umbilical cord (UC). These screenings examine several anatomical standard planes formally defined in guidelines~\cite{bhide2013isuog} developed by the International Society of Ultrasound in Obstetrics and Gynecology (ISUOG), which aims to ensure accurate measurement. In practice, conducting such examinations while adhering to the guidelines is technically demanding and often relies on the experience of the operators. However, due to the shortage of sonographers, third-trimester screenings are often performed by clinicians not specialized in ultrasound acquisition~\cite{edvardsson2016physicians,recker2021point}. To assist these clinicians in acquiring high quality images, this paper presents an automatic system for operator guidance in assessment of the umbilical artery (UA), which is a standard plane crucial for monitoring of the fetal blood supply, but has rarely been studied in existing works.

Examination of blood flow with ultrasound is performed over two steps: identifying and placing a measurement gate on a suitable site with color Doppler imaging, followed by measuring the blood flow curve in the form of spectral Doppler waveform with pulsed Doppler (see Fig.~\ref{fig:mca_ua}). These waveforms, which allow assessment of UA as well as the middle cerebral artery (MCA), play an important role in the third-trimester ultrasound screening~\cite{kennedy2019radiologist}. A number of machine learning techniques have been used to support analysis of Doppler waveforms. Multiple works~\cite{hoodbhoy2018machine,naftali2022novel} have explored the automatic diagnosis on waveform images acquired by clinicians. Hoodbhoy et al.~\cite{hoodbhoy2018machine} identified fetuses at increased risk of adverse perinatal outcomes with multiple kernel learning. Naftali et al.~\cite{naftali2022novel} built a support vector machine, a K-nearest model, and a logistic regression model for identifying unseen UC abnormalities from the waveforms. While these studies show a clear potential for AI-based diagnosis, the studies were conducted using waveforms acquired by experienced clinicians. As noted by Necas~\cite{necas2016obstetric}, even with the availability of guidelines, mistakes are still commonly observed in practice.

Meanwhile, acquisition of an optimal Doppler waveform requires consistent, manual adjustment of various parameters, which increases the mental load on the operator~\cite{wang2018deep}. This motivates research in automation of the ultrasound acquisition procedure. In this direction, MCANet~\cite{wang2018deep} is developed to propose positions for measurement gate placement. In contrast to MCANet, we consider Doppler measurement at the UA, and are, to the best of our knowledge, the first to address this problem. In addition, we also introduce a study protocol for assessing the quality of the acquired Doppler waveforms. In summary, we contribute with \textbf{1)} an automatic detection of appropriate sites for UA Doppler measurement, while taking both location and insonation angle into account, and \textbf{2)} an automatic evaluation of the Doppler waveforms to ensure sufficient quality.

\section{Method}

We propose an image analysis pipeline that is based directly on the ISUOG guidelines (see Table \ref{tab:isoug_guideline}). %\mx{maybe put them in a table?}\anders{That would take more space, no?}
The first two criteria relate to the presentation of UA in the chosen ultrasound plane, whereas the remaining three criteria relate to the quality of the resulting waveforms. As such, the ISUOG criteria align well with a 2-step process. First, we assist the ultrasound operator by giving feedback on the presentation of the UA and suggesting a suitable gate location for measuring the Doppler spectrum. Next, we consider the resulting spectrum, either accepting or rejecting the combination of ultrasound plane and gate. In this way, we ensure that the examination maintains a high quality, while also giving the operator feedback on potential problems in the image acquisition. 

% \begin{itemize}
%     \item Anatomical site - the Doppler signal must be measured in a free loop of the UC.
%     \item Angle of insonation - the angle to the blood flow should be less than 30 $^\circ$, indicating a near-vertical presentation of the blood vessel.
%     \item Image clarity - the resulting Doppler signal should be clear, without artifacts, and with an accurate trace.
%     \item Sweep speed - the resulting Doppler signal should have 3-10 acceptable waveforms visible.
%     \item Dynamic range - in the resulting Doppler signal, the waveforms should occupy more than 75\% of the y-axis.
% \end{itemize}

% \begin{table}
% \centering
% \caption{ISUOG guideline}
% \begin{tabular}{llll}
% Anatomical site     & the Doppler signal must be measured in a free loop of the UC &  &   \\
% Angle of  & the angle to the blood flow should be less than 30 $^\circ$,                            &  &   \\
% insonation & indicating a near-vertical presentation of the blood vessel & & \\
% Image clarity       & the resulting Doppler signal should be clear, without artifacts, and with an accurate trace                           &  &   \\
% Sweep speed         & the resulting Doppler signal should have 3-10 acceptable waveforms visible                           &  &   \\
% Dynamic Range       & in the resulting Doppler signal, the waveforms should occupy more than 75\% of the y-axis                           &  &  
% \end{tabular}
% \end{table}
\begin{table}[t]
\caption{ISUOG guidelines on using Doppler ultrasound in obstetric}
\centering
\label{tab:isoug_guideline}
\begin{tabular}{|l|l|}
\hline
\textbf{Criterion} &  \textbf{Description} \\
\hhline{|=|=|}
Anatomical site &  the Doppler signal must be measured in a free loop of the UC \\
Angle of insonation &  the angle to the blood flow should be less than $30^\circ$, indicating \\
 & a near-vertical presentation of the blood vessel \\
Image clarity & the resulting Doppler signal should be clear, without artifacts, \\
& and with an accurate trace \\
Sweep speed & resulting Doppler signal should have 3-10 waveforms visible \\
Dynamic range & in the resulting Doppler signal, the waveforms should occupy \\
& more than 75\% of the y-axis \\
\hline
\end{tabular}
\end{table}

\subsection{Deep Learning-based Processing of Color Doppler Images}
\label{sec:doppler_image_processing}
\begin{figure}[t]
    \centering
    \includegraphics[width=0.8\textwidth]{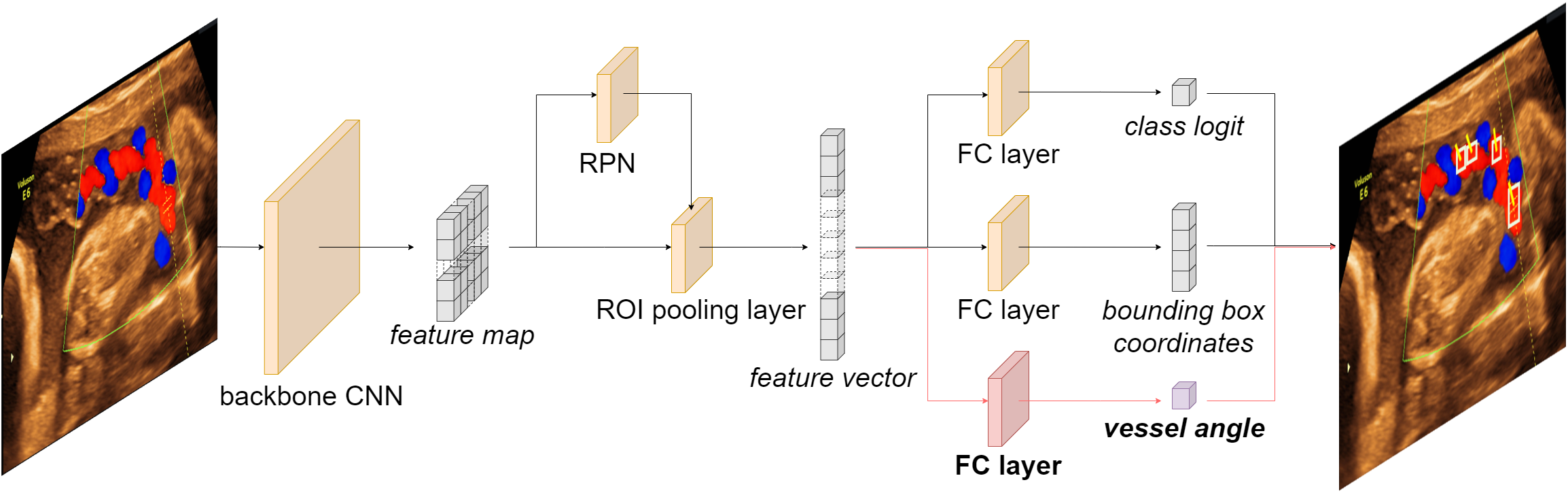}
    \caption{Network architecture: Faster R-CNN with an extra sibling FC layer branch added for vessel angle prediction.}
    \label{fig:network_architecture}
\end{figure}

As a first step in our pipeline, we need to identify a vessel, and a gate location within the vessel. This vessel needs to be an artery (as opposed to a vein) satisfying the first two ISUOG criteria. While it might seem trivial to segment vessels that are colored red and blue and measure their angle, Fig.~\ref{fig:mca_ua} illustrates how this problem is significantly harder for the UA, where the vessel twists and turns around itself, than it is for the MCA. We tackle this as a modified object detection problem where, instead of just detecting potential gate locations within vessels as being acceptable or not, we add a regression head to the detection algorithm in order to encode its angle (see Fig.~\ref{fig:network_architecture}).

\paragraph{Modifying Faster R-CNN's head to predict vessel angle at proposed locations.}
\label{sec:modified_fasterrcnn_to_pred_angle}

We base our UA gate location algorithm on the Faster R-CNN object detection algorithm \cite{ren2015faster}. Faster R-CNN first generates a feature map using a backbone CNN network, which is fed to a Region Proposal Network (RPN) to propose candidate regions of interest (ROIs). Then, an ROI pooling layer is employed to extract, pool, and flatten features corresponding to the proposed ROIs, which are finally passed to two sibling branches of fully connected (FC) layers for classification and bounding box regression of the proposed ROIs. The loss function proposed in the original paper is given by:
\begin{align}
    L_{FasterRCNN}(\{p_i\}, \{t_i\}) = \frac{1}{N_{cls}} \sum_i L_{cls}(p_i, p_i^*) + \lambda \frac{1}{N_{reg}}\sum_i p_i^* L_{reg}(t_i, t_i^*) \label{eq:faster_rcnn_loss_func}
\end{align}
where $p_i$ is the predicted probability that the $i$-th proposed ROI at the predicted coordinates $t_i$ is a valid object, with $p_i^*$ and $t_i^*$ as the corresponding ground-truths. The log loss $L_{cls}$ is normalized by the batch size $N_{cls}$, and the smooth L$_1$ loss $L_{reg}$ by the number of proposed ROIs $N_{reg}$, with $\lambda$ balancing the terms.

In our setting, we want to detect potential gate locations that satisfy both the anatomical site and angle requirements of the ISUOG criteria. To achieve this, the standard classification branch predicts whether the anatomical site criterion is fulfilled. Then, we enrich the Faster R-CNN model with an extra sibling FC layer branch to predict the underlying vessel’s angle at each proposed ROI.

\begin{figure}
    \centering
    \begin{subfigure}[t]{0.32\textwidth}
        \centering
        \includegraphics[width=\textwidth]{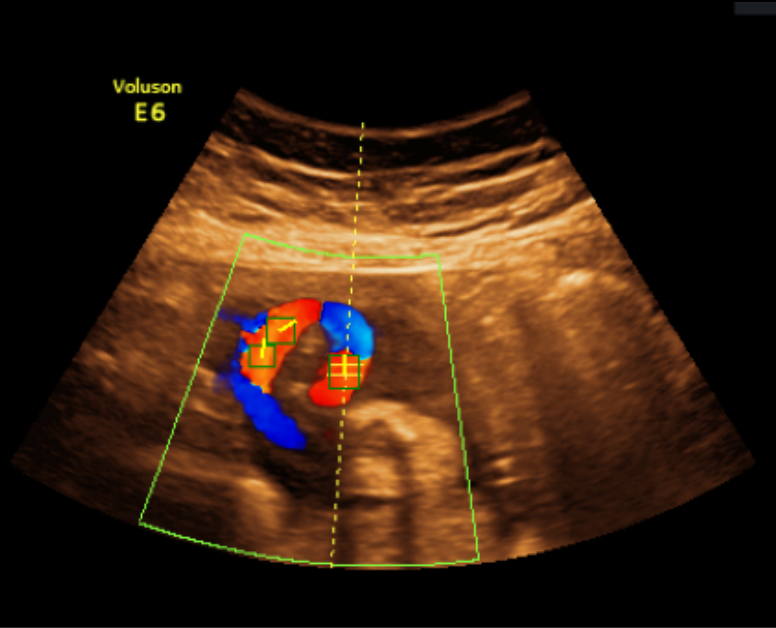}
        \caption{Expert annotated boxes and tangential lines}
        \label{fig:bbox_angle_gt}
    \end{subfigure}
    \hfill
    \begin{subfigure}[t]{0.32\textwidth}
        \centering
        \includegraphics[width=\textwidth]{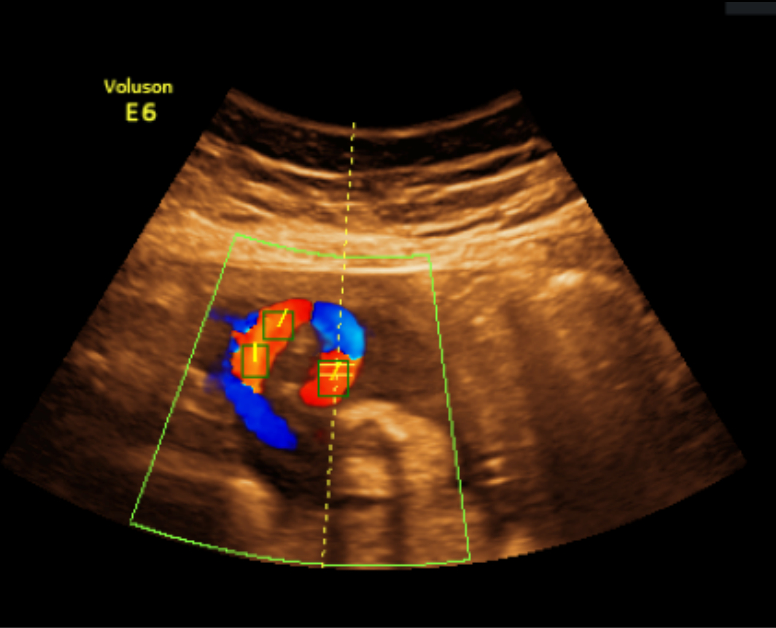}
        \caption{Model predicted boxes and tangential lines}
        \label{fig:bbox_angle_pred}
    \end{subfigure}
    \hfill
    \begin{subfigure}[t]{0.32\textwidth}
        \centering
        \includegraphics[width=\textwidth]{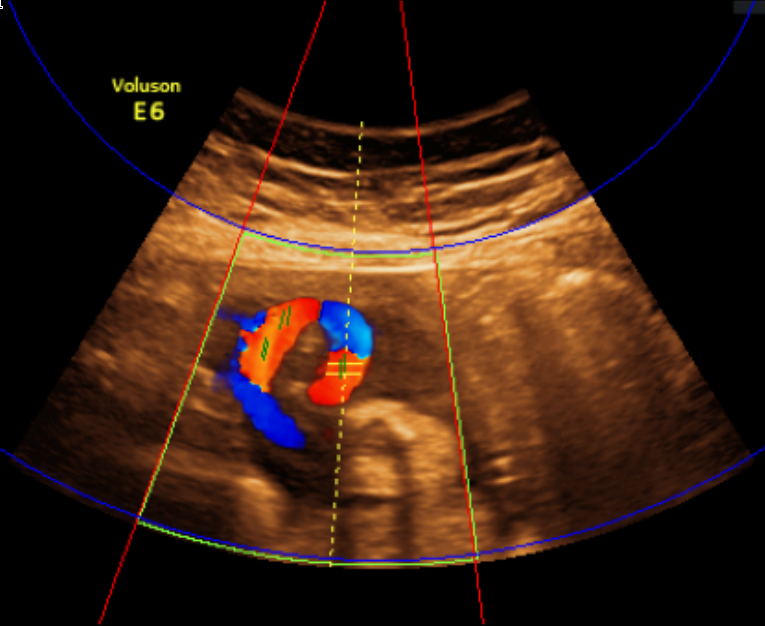}
        \caption{Vectors pointing in direction of ultrasound source}
        \label{fig:bbox_angle_probe_vectors}
    \end{subfigure}
    \caption{Different stages in processing Color Doppler image.}
    \label{fig:bbox_angle}
\end{figure}

To train this modified model, we included another term in Eq. \ref{eq:faster_rcnn_loss_func} to regress the predicted angle against ground truth annotations (see Fig.~\ref{fig:bbox_angle_gt}):
    \begin{align}
        L(\{p_i\}, \{t_i\}, \{a_i\}) = L_{FasterRCNN}(\{p_i\}, \{t_i\}) + \mu \frac{1}{N_{reg}} \sum_i p_i^* L_{reg}(a_i, a_i^*) \label{eq:modified_loss_func}
    \end{align}
where $a_i$ and $a_i^*$ are the predicted and ground truth angle of the $i$-th proposed ROI, and $\mu$ is another balancing factor. $\lambda$ and $\mu$ are set to 1 and 10, respectively. %\anders{As a minimum, we need to define the terms that goes into loss (2). Maybe OK we do not do it for (1).}

\paragraph{Locating source of ultrasound beam.}
\label{sec:locate_ultrasound_source}
The vessel angle estimated using our modified Faster-RCNN network indicates the vessel's direction (see Fig.~\ref{fig:bbox_angle_pred}). However, the angle of insonation needed to assess the second ISUOG criterion refers to the angle of the vessel with respect to the direction of the ultrasound beam. Hence, to assess whether the angle is acceptable and give suitable feedback to the operator, we need to determine the direction of the ultrasound beam and use it to calibrate the angle prediction from the network.

\begin{figure}[t]
    \centering
    \begin{subfigure}[t]{0.49\textwidth}
        \centering
        \includegraphics[width=\textwidth]{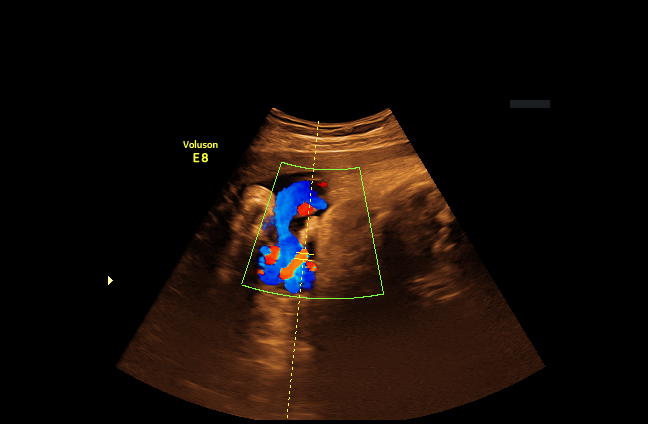}
        \caption{Green Doppler box}
        \label{fig:green_doppler_box}
    \end{subfigure}
    \hfill
    \begin{subfigure}[t]{0.49\textwidth}
        \centering
        \includegraphics[width=\textwidth]{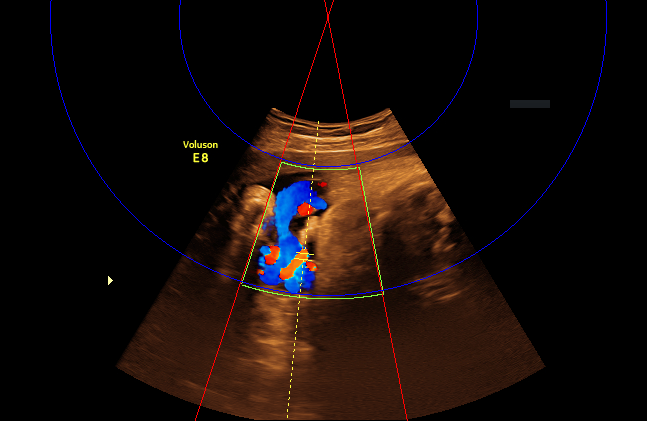}
        \caption{Detected line (red) and arc (blue)}
        \label{fig:green_doppler_box_detection_overlay}
    \end{subfigure}     
    \caption{Locating source of the ultrasound beam by using the green Doppler box.}
    \label{fig:probe_direction_detection}
\end{figure}

We begin with determining the direction of the ultrasound beam by utilizing the green Doppler box, which becomes active during color Doppler acquisition (see Fig. \ref{fig:green_doppler_box}). A binary mask of the green Doppler box is first obtained by color thresholding on the RGB pixels, followed by enhancing the mask using a watershed transform~\cite{huang2004watershed}. Next, a Hough line detection algorithm~\cite{aggarwal2006line} is used to identify the two radial line segments on each side of the box. The ultrasound source is located at the intersection between the two line segments. As an extra step to prevent erroneous detection, we verify that we are able to detect the two arc lines with Hough circle detection when the center location is constrained to be at the intersection point from the previous two Hough lines (see Fig. \ref{fig:green_doppler_box_detection_overlay}).

\paragraph{Determining the angle of insonation.}
Next, from the identified location of the ultrasound source, it is straightforward to compute a vector map pointing at the direction of the ultrasound source from each pixel (see Fig.~\ref{fig:bbox_angle_probe_vectors}). By expressing the vessel angle predicted by the network as another vector, the angle of insonation can be easily calculated using the law of cosine.

Based on the above, we obtain a detection algorithm that proposes, to the operator, gate locations fulfilling both anatomical and angle requirements. This algorithm is developed using images acquired with different models of GE scanners. Adjustments for cross-vendor applications is likely necessary, but we expect the method to work equally well, since it is a convention in Doppler ultrasound to use variations of red and blue while encoding flow directions~\cite{bhide2013isuog}.

%\begin{itemize}
%    \item 
%    \item 
    % \begin{align}
    %     \theta = \arccos{\frac{v_{vessel} \cdot v_{probe}}{|v_{vessel}| \cdot |v_{probe}| }}
    % \end{align}
%\end{itemize}

\subsection{Processing of Pulsed Doppler Spectrum Images}
\label{sec:spectrum_processing_pipeline}

From the suggested gate locations (see Sec.~\ref{sec:modified_fasterrcnn_to_pred_angle}), an operator could choose the best one for spectrum acquisition with pulsed Doppler. Next, we present our processing to assess the quality of the resulting spectrum.
% \aasa{Include this in the appropriate place: We leveraged heavily on functions implemented in the Scipy and Scikit-image package.}

\paragraph{Segmentation of the Doppler Spectrum.}
\label{sec:segmentation_doppler_spectrum}
Prior to analyzing the Doppler spectrum, all irrelevant visual confounders, such as tracing lines indicating measurements made by the scanner (see Fig. \ref{fig:mca_ua}), are first removed from the image to prevent bias in our method. Since the confounders are all colored, they can be delineated from the grey spectrum by identifying pixels with large variance across the color (i.e. RGB) channel, and removed by inpainting with biharmonic functions.

Next, the spectrum is segmented using a random walk segmentation algorithm~\cite{grady2006randomwalk}, as watershed is found to be insufficient. From the mask of the segmented spectrum, the envelope of the spectrum can be identified from the boundary. Then, the $x$-axis line of the spectrum is detected using the Hough line detection algorithm, and used to remove anything below and fill any holes above. %See supplementary material for examples and more details.

\paragraph{Identification of Individual Waveforms.}
From the spectrum segmentation, peaks and valleys are detected along the upper envelope. To avoid over-detection, the envelope is Gaussian-smoothed prior to running the detection algorithm. Besides that, the detected peaks (and valleys) are forced to have a minimum distance of 70 pixels between each other. Individual waveforms are then identified with the rule that a complete waveform starts and ends at valleys with a peak in between.

\begin{figure}[t]
    \centering
    % \begin{subfigure}[t]{0.8\textwidth}
    %     \centering
    %     \includegraphics[width=\textwidth]{spectrum_raw}
    %     \caption{Spectrum image with visual confounders}
    %     \label{fig:spectrum_raw}
    % \end{subfigure}
    % \hfill
    % \begin{subfigure}[t]{0.8\textwidth}
    %     \centering
    %     \includegraphics[width=\textwidth]{spectrum_detected_peak_valley}
    %     \caption{Detected peaks and valleys}
    %     \label{fig:spectrum_detected_peak_valley}
    % \end{subfigure}
    % \hfill
    % \begin{subfigure}[t]{0.8\textwidth}
    %     \centering
    %     \includegraphics[width=\textwidth]{labelled_waveform}
    %     \caption{Identified individual waveforms}
    %     \label{fig:labelled_waveform}
    % \end{subfigure}
    % \hfill
    % \begin{subfigure}[t]{0.8\textwidth}
    %     \centering
    %     \includegraphics[width=\textwidth]{waveform_clarity}
    %     \caption{Image clarity. Green is good (mean intensity > 0.56), yellow is moderate, red (mean intensity < 0.36) is poor.}
    %     \label{fig:waveform_clarity}
    % \end{subfigure}
    % \hfill
    % \begin{subfigure}[t]{0.8\textwidth}
    %     \centering
    %     \includegraphics[width=\textwidth]{waveform_clarity_height}
    %     \caption{Individual waveforms identified with overlay information of their clarity and height. Green is good clarity (mean intensity > 0.56), yellow is moderate, red (mean intensity < 0.36) is poor. Dotted line means height is less than 75\%. At least one waveform > 75\% is sufficient.}
    %     \label{fig:waveform_clarity_height}
    % \end{subfigure}
    % \caption{\ck{to fill}}
    % \label{fig:spectrum_processing}
    \includegraphics[width=0.8\textwidth]{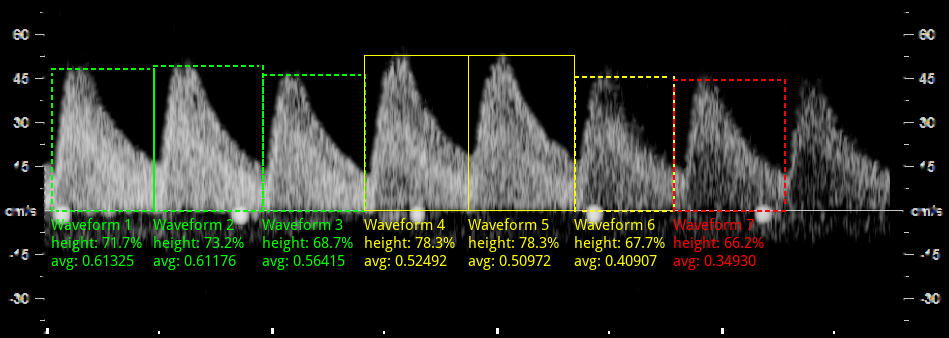}
    \caption{Waveform clarity and height. Good clarity is indicated in green (mean intensity > 0.56), moderate in yellow, and poor in red (mean intensity < 0.36) is poor. Dotted line indicates waveforms with height less than 75\%.}
    \label{fig:waveform_clarity_and_height}
\end{figure}

\paragraph{Determining Waveform Properties.}
Having identified individual waveforms, we need to quantify their quality in order to assess whether the waveform as a whole satisfies the last three ISUOG criteria. To obtain this, the spectrum image pixels are first rescaled to floating point values between 0 and 1. Using the spectrum segmentation, we mask out the relevant part of the spectrum to compute the mean intensity of pixels in each individual waveform. Two threshold values are defined to classify the waveform as having good, moderate, or poor clarity (see Fig. \ref{fig:waveform_clarity_and_height}). In our experiment, these thresholds are set empirically to 0.56 and 0.36. Heights of the waveform are determined from the peak coordinate and the $x$-axis line, followed by expressing them as a percentage of the positive $y$-axis.

\section{Dataset}
%\textit{Chun Kit to expand; Zahra/Alberto to help}
\label{sec:dataset}
\begin{comment}
\begin{itemize}
    \item 657 DICOM images of umbilical artery flow measurement were randomly sampled from a private dataset
    \item From the DICOM images, two regions were cropped and exported:
    \begin{itemize}
        \item the ultrasound image together with the color Doppler overlays
        \item the power Doppler spectrum
    \end{itemize}
    \item The ultrasound images were subsequently annotated by a clinical expert with LabelMe.
    \item Within each image, correct anatomical sites (i.e. the umbilical artery) were annotated as completely as possible with bounding boxes.
    \item Additionally, angle of the blood vessel beneath every bounding box was annotated with a tangential line.
    \item Note that due to the inherent ambiguity in identifying the umbilical artery from an ultrasound image, we are unable to claim that we've annotated all anatomical sites completely.
\end{itemize}
\end{comment}

The data used for this study were gathered under permission (ANONYMOUS) from a national screening database containing images acquired between 2009 and 2019 as part of the standard prenatal care. In total, 657 DICOM images of Doppler ultrasound examination at the UA were retrieved from the database. From each image, two regions were cropped and exported separately: (1) the b-mode ultrasound image with color Doppler overlays (for simplicity, we will call this the color Doppler image), and (2) the pulsed Doppler spectrum image.

% On each image, regions corresponding to the correct anatomical site (i.e. the UA) were sparsely annotated with bounding box.
Subsequently, the color Doppler images were annotated by a clinician using the open-source tool LabelMe \cite{wada2018labelme}. On each image, regions corresponding to the correct anatomical site (i.e. the UA) were annotated with bounding box, which was done sparsely due to the inherent difficulty in annotating all correct anatomical sites with just bounding boxes. Finally, a tangential line going in the direction of the blood vessel were added on top of each annotated bounding box to indicate the vessel's angle (see Fig. \ref{fig:bbox_angle_gt}).

% Note that due to the inherent ambiguity in identifying the umbilical artery from an ultrasound image, we are unable to claim that we've annotated all anatomical sites completely.

% The data used for this study were gathered under permission (ANONYMOUS)
% from a national screening database containing images acquired between 2009 and
% 2019 as part of the standard prenatal care. Images were retrieved that present
% both the Doppler ultrasound scan and the relative Doppler spectrum. These
% were cropped and used separately, for the two stages in the pipeline.%\anders{Do we want to mention the 600.000? It sounds so impressive, that it would immediately invite questions from the reviewers about, why we then only use 657?}   

% From the main database 657 images were extracted and the Doppler ultrasound scan region was annotated by the medical expert using the open-source tool LabelMe\footnote{\url{https://github.com/wkentaro/labelme}}. In each image, the expert annotated potential gate locations within the Umbilical Artery using bounding boxes. This gives a sparse annotation as for each section of UA only a few gate locations were annotated. For each bounding box, the expert annotated the angle of the blood vessel using a tangential line going in the direction of the vessel in that specific section.

% It has to be considered that due to the inherent ambiguity in differentiating the umbilical artery from the umbilical vein in a still ultrasound image, we are unable to claim that we have annotated all anatomical sites completely.

\section{Experiments and Results}
\subsection{Experimental Settings} 
The modified Faster R-CNN model was implemented in PyTorch 1.13.1 on an AlmaLinux 8.7 system. The model backbone is a ResNet50 FPN pre-trained on COCO~\cite{lin2014microsoft} and fine-tuned on our annotated images at 80:20 train-test split with SGD optimizer, which converged after 20 epochs. The learning rate was 0.005 at the first epoch and reduced 0.1 times every 3 epochs. The batch size was set to 2. This training took less than 1 hour on an NVIDIA RTX A6000 GPU. The experiments on waveform quality assessment were conducted on a server (CPU: EPYC 7252 8-Core Processor; RAM: 32GB), using image processing algorithms implemented in Scipy 1.10.0 and Scikit-Image 0.19.3.

\subsection{Results}
\label{sec:results}
%We achieved $AP_{50}$ and $AR_{50}$ of 36.7\% and 59.1\% respectively on our test set, leading to our F1-score of 45.28\%.
% Since it is challenging to annotate all correct anatomical sites with just bounding boxes (see Sec. \ref{sec:dataset}), the ground-truth labels in our dataset are under-annotated.
\paragraph{Automatic gate placement.} The ground-truth gate locations in our dataset are under-annotated (see Sec. \ref{sec:dataset}). Hence, to better estimate the error in our angle prediction, we compared each model-suggested gate position box with the closest ground truth box, matched according to the Euclidean distance between the centroid of the two boxes. Pairs separated with a too-far distance (greater than 10 pixels; see \ref{sec:appendix_model_validation}) were excluded from this analysis. With a 5-fold cross validation, $85.37\%$ (std $2.97\%$) of the ground truth has a matched prediction, with a mean absolute error of $19.36^\circ$ (std $0.47^\circ$) in angle prediction.

Additionally, an experienced clinician was engaged to check through the model predictions on 737 test images and reject images with at least one unacceptable box prediction, giving a conservative performance estimate. This gave an acceptance rate of $68.1\%$.

%we calculated the mean absolute error to be 0.293 radians when the IoU threshold is 0.50. \ck{this isn't the most exciting result; are we able to somehow justify it?} \aasa{where does the IoU come in here? Is it used to assess whether or not to keep boxes?} Our low precision is rather expected. Since it's unlikely that we've annotated our dataset completely (see Sec. \ref{sec:dataset}), we are unable to accurately estimate the false positive rate. Nonetheless, this prove-of-concept demonstrates that our model can provide guidance to the human operator on where are the correct locations to make the Doppler measurement.

\begin{figure}[t]
    \centering
    \begin{subfigure}[t]{0.48\textwidth}
        \centering
        \includegraphics[width=\textwidth]{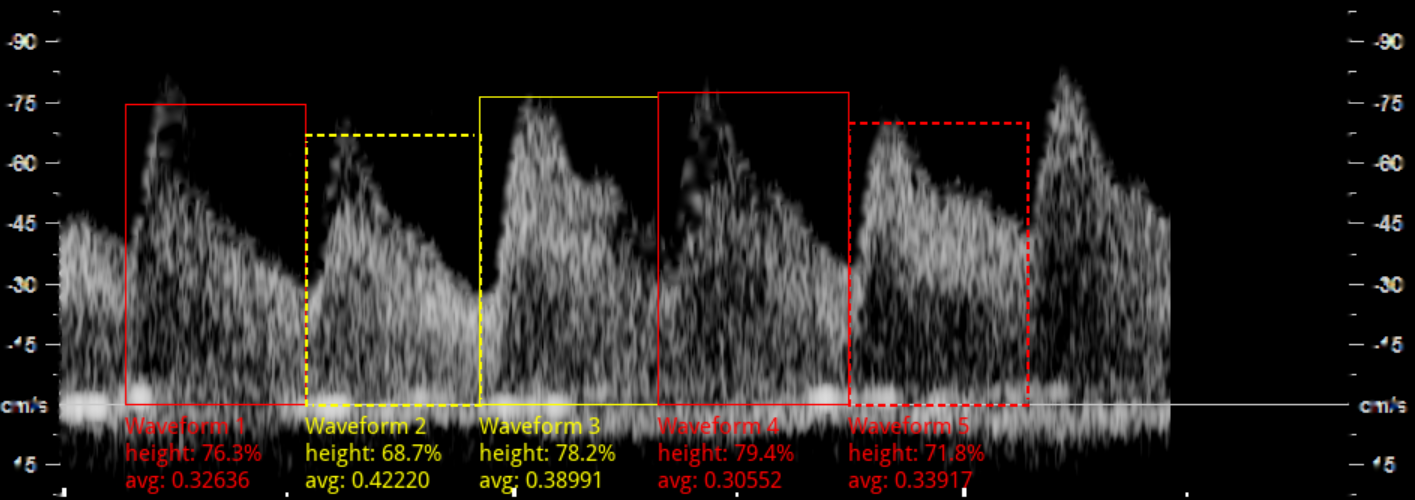}
        \caption{Less than 3 consecutive waveforms with at least moderate clarity}
        \label{fig:failed_waveform_example_04}
    \end{subfigure}
    \hfill
    \begin{subfigure}[t]{0.48\textwidth}
        \centering
        \includegraphics[width=\textwidth]{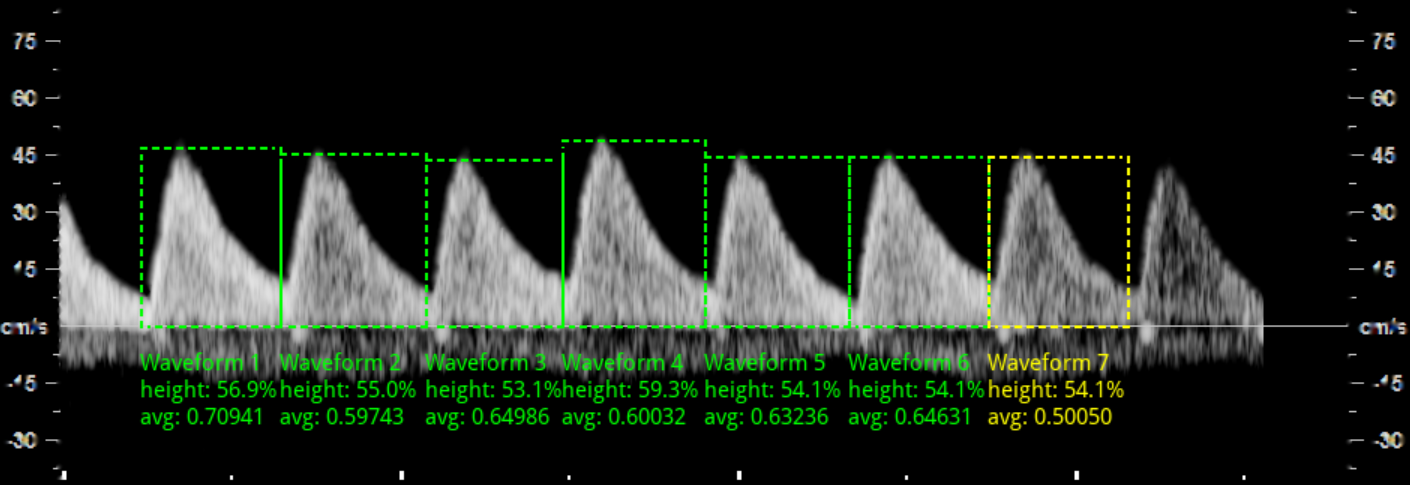}
        \caption{Waveforms not occupying 75\% of the y-axis}
        \label{fig:failed_waveform_example_05}
    \end{subfigure}
    % \hfill
    % \begin{subfigure}[t]{0.8\textwidth}
    %     \centering
    %     \includegraphics[width=\textwidth]{failed_waveform_example_03}
    %     \caption{Failed example 3}
    %     \label{fig:failed_waveform_example_03}
    % \end{subfigure}
    \caption{Example spectrums not meeting the ISUOG criteria. Zoom in for details.}
    \label{fig:failed_waveform_examples}
\end{figure}

\paragraph{Waveform quality assessment.}
% We evaluated the waveform quality assessment performance of our method on the 657 spectrum images extracted previously\anders{More concrete: How did we evaluate them?}nd
We applied the proposed waveform quality assessment pipeline on the 657 spectrum images. Fig.~\ref{fig:failed_waveform_examples} presents two images that has failed to meet the ISUOG waveform clarity and dynamic range criteria, identified by our method. This demonstrates the promising potential of our pipeline being alternatively deployed as a retrospective waveform quality assessment tool.

% \subsection{Parameter tuning for spectrum processing}
% \textit{Chun Kit to write}
\section{Discussion and Conclusion}
We attempted to address a real clinical problem in this paper: providing guidance to an operator on the Doppler assessment of UA, which is significant due to the shortage of skilled sonographers. We proposed a method for guiding the operator on where to place the measurement gate, which has shown promising performance on our test data. Furthermore, another method was proposed to assess the quality of the acquired spectrum, designed to be well-aligned to the criteria specified in the ISUOG guidelines. This provides a comprehensive assessment of the scan quality, ensuring compliance to criteria such as "sweep speed" and "dynamic range", which often get neglected in actual clinical practice.

% The rule-based assessment of the waveform might seem somewhat redundant given that newer scanners outline the waveform, and if they cannot the operator knows that the quality is too poor. However, we do not only consider the waveform clarity, but also the other ISUOG criteria - "sweep speed" and "dynamic range" - which tend to be neglected in clinical practice. We note that our rule-based method can be easily adapted by altering the parameters, which allows users to encode their own standards for quality control.

In the future, we aim to extend the method for processing live video stream directly. While new challenges is foreseen with the incorporation of temporal information, better performance can be expected in multiple ways, such as an improved accuracy in differentiating between the pulsating artery and the non-pulsating vein. Overall, we have demonstrated the potential of our method in resolving the clinical problem that has motivated its development.

% \paragraph{mean average precision of the bounding boxes and L1/L2 distance on angles}

% \paragraph{QA result: how many of the DICOM images are ok/nok}

% \paragraph{other discussions}
\clearpage

\bibliographystyle{splncs04}
\bibliography{mybibliography}

\clearpage

\pagenumbering{roman}
\setcounter{figure}{0}

\appendix

\renewcommand{\thesection}{Appendix \arabic{section}}
\renewcommand{\thefigure}{\Roman{figure}}
\renewcommand{\thesubfigure}{\Alph{subfigure}}

\section{Evaluating Model's Performance}
\label{sec:appendix_model_validation}
\begin{figure}
    \centering
    \begin{subfigure}[t]{0.48\textwidth}
        \centering
        \includegraphics[width=\textwidth]{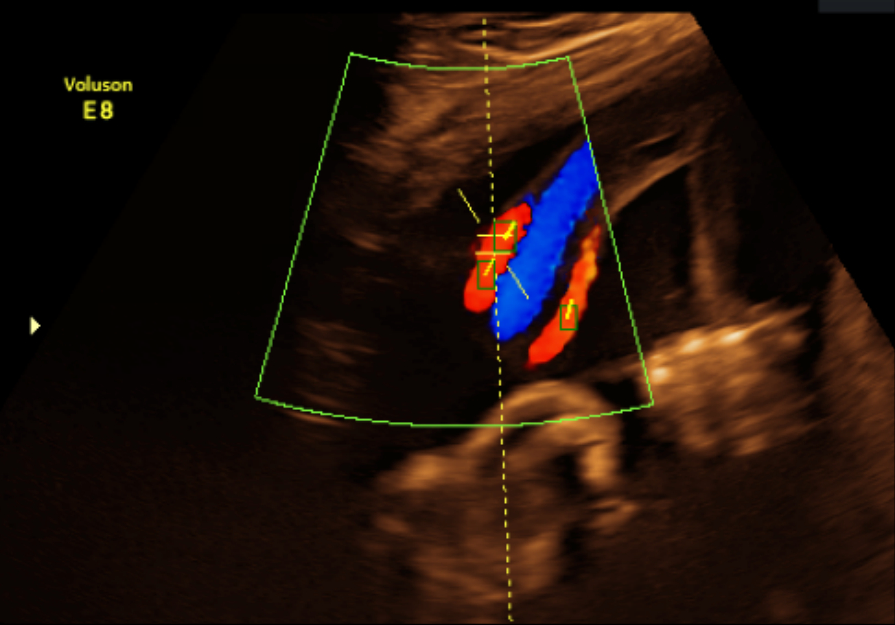}
        \caption{ground truth}
        \label{fig:model_validation_gt}
    \end{subfigure}
    \hfill
    \begin{subfigure}[t]{0.48\textwidth}
        \centering
        \includegraphics[width=\textwidth]{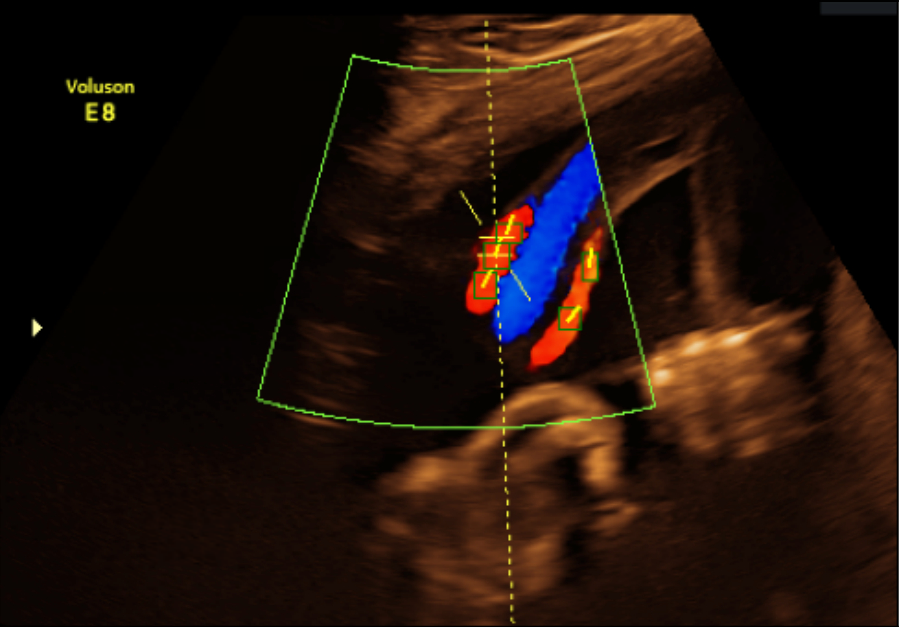}
        \caption{prediction}
        \label{fig:model_validation_pred}
    \end{subfigure}
    \caption{Manually-annotated (A) versus model-predicted (B) bounding boxes and angles. The correct anatomical sites are sparsely annotated due to laborious nature of the task, which means not every correct anatomical sites are annotated in our ground truth dataset.}
    \label{fig:model_validation}
\end{figure}
\begin{figure}
    \centering
    \begin{subfigure}[t]{0.77\textwidth}
        \centering
        \includegraphics[width=\textwidth]{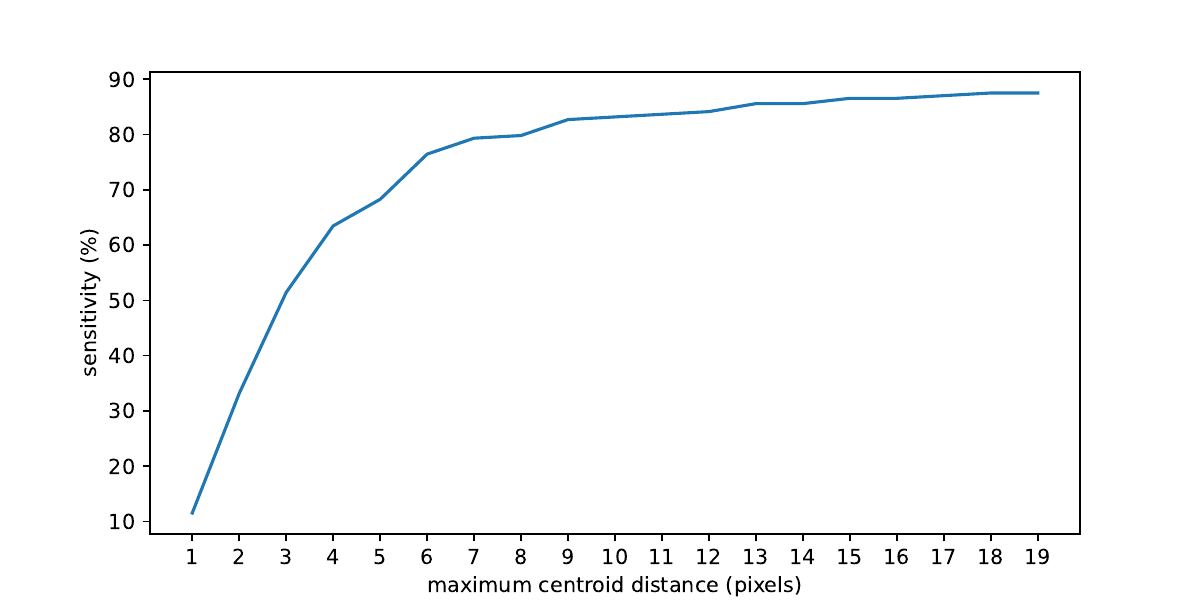}
        \caption{$sensitivity$ against $n$ (the higher the better)}
        \label{fig:validation_sensitivity}
    \end{subfigure}
    \hfill
    \begin{subfigure}[t]{0.77\textwidth}
        \centering
        \includegraphics[width=\textwidth]{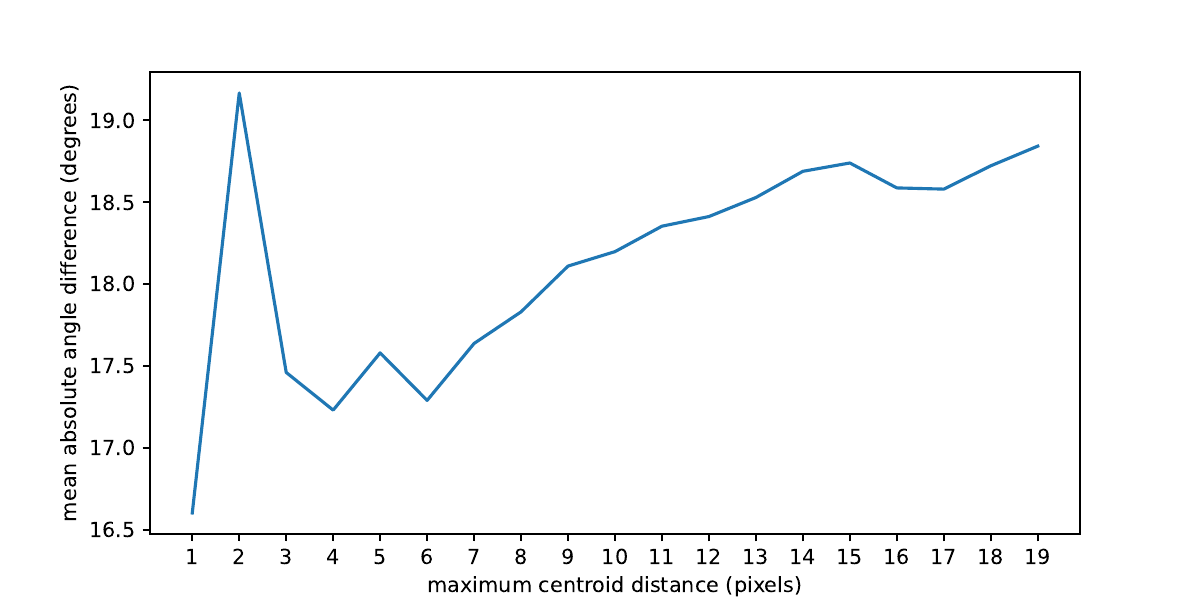}
        \caption{$\Bar{L}_{1}$ against $n$ (the lower the better)}
        \label{fig:validation_mean_angle_err}
    \end{subfigure}
    \caption{Evaluation of model's performance}
    \label{fig:validation_result}
\end{figure}

Since the correct anatomical sites are under-annotated in our dataset (see Fig. \ref{fig:model_validation}), we are unable to determine the false-positive in our predictions and, subsequently, the precision of our model. Hence, we chose to evaluate our model performance with a sensitivity-based approach. For every ground-truth box (see Fig. \ref{fig:model_validation_gt}), we identify the corresponding nearest predicted box (see Fig. \ref{fig:model_validation_pred}) by determining box pair that yields the shortest Euclidean distance:
% \begin{equation}
%     \mathbf{t^{*}_{i}} \xleftrightarrow{paired} \mathbf{t_{\tilde{j}}}, \quad \tilde{j} = \argmin_{j} \lVert \mathbf{t^{*}_{i}} - \mathbf{t_{j}} \rVert_{2} \nonumber
% \end{equation}
\begin{equation}
    f(\mathbf{t^{*}_{i}}) = \mathbf{t_{\tilde{j}}}, \quad \tilde{j} = \argmin_{j} \lVert \mathbf{t^{*}_{i}} - \mathbf{t_{j}} \rVert_{2} \nonumber
\end{equation}
% \begin{equation}
% f(\mathbf{t^{*}_{i}}) = 
%     \begin{cases}
%         \mathbf{t_{\tilde{j}}}, \quad \tilde{j} = \argmin_{j} \lVert \mathbf{t^{*}_{i}} - \mathbf{t_{j}} \rVert_{2}, & \text{if } x\geq 1\\
%         0,              & \text{otherwise}
%     \end{cases} \nonumber
% \end{equation}
where $\mathbf{t^{*}_{i}}$ represents centroid of the $i$-th ground-truth box, $\mathbf{t_{j}}$ centroid of the $j$-th predicted box, $\tilde{j}$ the index of the nearest predicted box, and $f$ the mapping function. Then, for a given threshold $n$, we say the ground truth box is successfully detected if its nearest predicted box is less $n$ pixels away:
\begin{align}
{T} &\coloneqq \{ \, \mathbf{t^{*}_{i}} \, \} \nonumber \\
\widetilde{T} &\coloneqq
\begin{cases}
\{ \, \mathbf{t^{*}_{i}} \mid \lVert \mathbf{t^{*}_{i}} - \mathbf{t_{\tilde{j}}} \rVert_{2} < n \, \}, & \text{if } \{ \, \mathbf{t_{j}} \, \} \neq \varnothing \\
\varnothing,              & \text{otherwise}
\end{cases} \nonumber
\end{align}
Finally, the sensitivity of our model is derived by calculating the percentage of successfully detected ground truth box:
\begin{align}
sensitivity &= \frac{|\widetilde{T}|}{|T|} \times 100\% \nonumber
\end{align}
while the inaccuracy in angle prediction is estimated using the mean $L_{1}$ distance between the ground truth angles $\{\,a^*_{i}\,\}$ and the predicted angles $\{\,a_{j}\,\}$:
\begin{align}
    \Delta A &= \{ \, \lVert a^*_{i} - a_{\tilde{j}} \rVert \, \} \nonumber \\
    \Bar{L}_{1} &= \frac{\sum \Delta A}{|\Delta A|} \nonumber
\end{align}
As expected, both the sensitivity and the angle inaccuracy increased with a larger threshold $n$ (see Fig. \ref{fig:validation_result}). Since a higher sensitivity and lower angle inaccuracy is desired, we reported the values with $n$ fixed to 10 (see Sec. \ref{sec:results} of the main article) to avoid over-glorifying our model in either of the two performance metrics.

\end{document}